\documentclass[sigconf]{acmart}

\usepackage[htt]{hyphenat}  

\usepackage[ruled,vlined,linesnumbered]{algorithm2e}

\usepackage{array}
\newcolumntype{P}[1]{>{\centering\arraybackslash}p{#1}}
\newcolumntype{R}[1]{>{\raggedleft\arraybackslash}p{#1}}

\usepackage{listings}
\usepackage{color}
\usepackage{multirow}
\usepackage{tcolorbox}  
\usepackage{balance}
\usepackage{url}
\usepackage{caption}
\usepackage{amsmath}
\usepackage{amsthm}

\usepackage{xcolor}
\usepackage{enumitem}

\usepackage{booktabs}
\renewcommand{\arraystretch}{0.9}

\definecolor{dkgreen}{rgb}{0,0.6,0}
\definecolor{gray}{rgb}{0.5,0.5,0.5}
\definecolor{mauve}{rgb}{0.58,0,0.82}
\definecolor{darkblue}{rgb}{0.0,0.0,0.6}
\definecolor{lightblue}{rgb}{0.0,0.0,0.9}
\definecolor{cyan}{rgb}{0.0,0.6,0.6}
\definecolor{darkred}{rgb}{0.6,0.0,0.0}

\lstdefinelanguage{XML}
{
  morestring=[s][\color{mauve}]{"}{"},
  morestring=[s][\color{black}]{>}{<},
  morecomment=[s]{<?}{?>},
  morecomment=[s][\color{dkgreen}]{<!--}{-->},
  stringstyle=\color{black},
  identifierstyle=\color{lightblue},
  keywordstyle=\color{red},
  morekeywords={android, name, resource, xmlns,xsi,noNamespaceSchemaLocation,type,id,x,y,source,target,version,tool,transRef,roleRef,objective,eventually}
}

\lstdefinestyle{codestyle}{
    backgroundcolor=\color{white},   
    commentstyle=\color{green!50!black},
    keywordstyle=\color{blue},
    numberstyle=\tiny\color{black},
    stringstyle=\color{red},
    basicstyle=\ttfamily\small,
    breakatwhitespace=false,         
    breaklines=true,                 
    captionpos=b,                    
    keepspaces=true,                 
    numbers=left,                    
    numbersep=5pt,                  
    showspaces=false,                
    showstringspaces=false,
    showtabs=false,                  
    tabsize=2,
    frame=none,
    xleftmargin=0pt,
    xrightmargin=0pt,
    moredelim=**[is][\color{white}\colorbox{gray!50}]{@}{@}, 
}

\copyrightyear{2026}
\acmYear{2026}
\setcopyright{cc}
\setcctype{by}
\acmConference[AST '26]{7th ACM/IEEE International Conference on Automation of Software Test (AST 2026)}{April 13--14, 2026}{Rio de Janeiro, Brazil}
\acmBooktitle{7th ACM/IEEE International Conference on Automation of Software Test (AST 2026) (AST '26), April 13--14, 2026, Rio de Janeiro, Brazil}
\acmPrice{}
\acmDOI{10.1145/3793654.3793745}
\acmISBN{979-8-4007-2476-3/2026/04}

\begin{document}

\title{Understanding and Detecting Platform-Specific Violations in Android Auto Apps}


\author{Moshood A. Fakorede}
\orcid{0009-0003-2057-9865}
\affiliation{%
  \institution{Louisiana State University}
  \city{Baton Rouge}
  \state{LA}
  \country{USA}}
\email{mfakor1@lsu.edu}

\author{Umar Farooq}
\orcid{0000-0001-7229-9847}
\affiliation{%
  \institution{Louisiana State University}
  \city{Baton Rouge}
  \state{LA}
  \country{USA}}
\email{ufarooq@lsu.edu}


\newcommand{\stitle}[1]
{\noindent\textup{\textbf{#1}}}
\newcommand\myNum[1]{\emph{{(#1)}}}



\begin{abstract}

Despite over 3.5 million Android apps and 200+ million Android Auto-compatible vehicles, only a few hundred apps support Android Auto due to platform-specific compliance requirements. Android Auto mandates service-based architectures in which the vehicle system invokes app callbacks to render the UI and handle interactions, which is fundamentally different from standard Activity-based Android development. Through an empirical study analysis of $98$ issues across $14$ Android Auto app
repositories, we identified three major compliance failure categories: media playback errors, UI rendering issues, and voice command integration failures in line with mandatory requirements for integrating Android Auto support. We introduce {\textsc{AutoComply}}, a static analysis framework capable of detecting these compliance violations through the specialized analysis of platform-specific requirements. {\textsc{AutoComply}}~constructs a Car-Control Flow Graph (CCFG) extending traditional control flow analysis to model the service-based architecture of Android Auto apps. Evaluating {\textsc{AutoComply}}~on $31$ large-scale open-source apps, it detected $27$ violations
(13$\times$ more than Android Lint), while no false positives were observed, achieving 2$\times$ faster analysis. Developers have acknowledged $14$ of these violations with $8$ fixes already implemented, validating {\textsc{AutoComply}}'s practical effectiveness.

\end{abstract}



\maketitle

\section{Introduction}
\label{sec:intro}
Mobile apps play a key role in communication, entertainment, and navigation, extending their influence into the automotive sector through platforms such as Android Auto. 
Android Auto enables drivers to access mobile apps safely via in-car infotainment systems. 
Although the Google Play Store lists over 3.5 million apps~\cite{number-of-apps-on-stores} and more than 200 million vehicles support Android Auto as of 2024~\cite{android-auto-number-of-cars}, only a small number of apps meet the platform's compliance requirements. 
As of March 2025, just 363 apps are Android Auto compliant. 
This disparity points to technical and design challenges that developers face when making their apps compliant for automotive use.
Addressing these challenges and understanding the specific compliance barriers is critical for expanding the availability of high-quality, in-vehicle app experiences.

\stitle{Android Auto platform requirements.}
In contrast to standard Android apps, Auto apps are built around a service-based architecture. 
In this model, background services serve as the primary components for handling user interaction, media browsing, and playback, rather than relying on activities.

To be considered compliant with the Android Auto platform, Android apps must meet several additional criteria defined by the Android Auto framework~\cite{androidauto}. 
An app is considered \emph{Android Auto compliant} if it satisfies the following requirements:
\emph{(1) Discoverability:} The app must be properly declared in the manifest with appropriate automotive metadata and service registrations to ensure it is discoverable by the system.
\emph{(2) Voice Integration:} The app must integrate required voice command APIs to support safe, hands-free interaction.
\emph{(3) UI Compliance:} The app must use designated UI templates and construct its content hierarchy through \texttt{MediaBrowserService}, ensuring that the interface renders and behaves correctly on in-car displays.
\emph{(4) Functional Requirements:} The app must implement category-specific callbacks and other platform-defined behaviors, particularly for media playback functionality.

\begin{figure}
    \centering
    \includegraphics[width=0.96\linewidth]{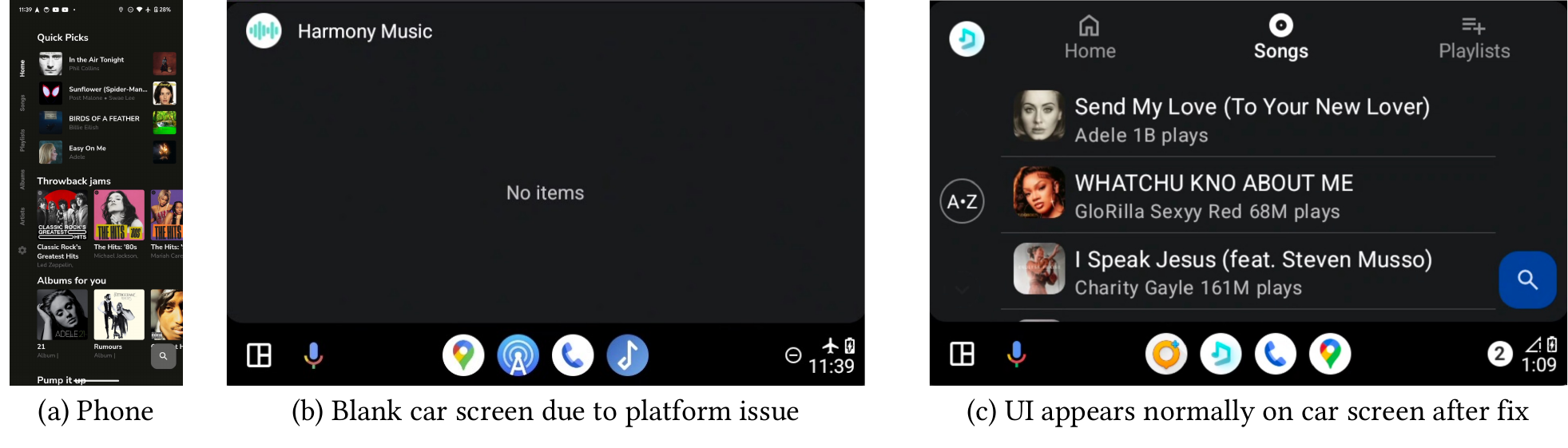}
    \vspace{-1.5em}
    \caption{Example Android Auto platform issue in an app, (a) app works fine on the phone. (b) Due to compliance issues with Android Auto, the app does not appear as intended on the car screen. (c) After we reported the issue to the developer, the developer was able to fix the problem.}
    \vspace{-1em}
    \label{fig:ui-issue}
    \vspace{-10pt}
\end{figure}

Figure~\ref{fig:ui-issue} illustrates a typical Android auto specific we discovered in which the app fails to meet some of these criteria. While the app displays its music library as intended on the phone (a), it fails to render properly on the car screen due to missing callbacks, preventing content rendering in Android Auto (b). After we identified and reported the missing callbacks, the developer implemented them, resolving the issue (c).

This example demonstrates how Android Auto's service-based architecture imposes requirements beyond standard Android development requirements. Such platform-specific requirements are difficult to detect manually and are often missed by existing static analysis tools, motivating our work.

\stitle{Formative Study.} 
Meeting Android Auto's above-mentioned compliance requirements poses technical and procedural challenges for developers. 
To better understand these barriers in practice, we conducted a formative study to identify the most frequent and impactful sources of compliance failure and to inform the development of automated methods for detecting and addressing these issues.

To conduct this study, we focus on open-source apps from F-Droid~\cite{fdroid} by collecting $4,387$ apps in total. 
We filter for those with Android Auto tags declared in their manifest files (i.e., \texttt{AndroidManifest.xml}), identifying $37$ apps with developers' intent to add Auto support. 
We subsequently analyzed the repositories of these apps, identifying $98$ Android Auto specific issues across $14$ repositories. 
Our analysis reveals that these issues fall primarily into three categories: \myNum{i} UI compliance issues (as shown in Figure~\ref{fig:ui-issue}), \myNum{ii} media playback challenges, and \myNum{iii} failure to implement voice command integration.
This comprehensive study provided valuable insights into the technical challenges developers face when adapting their apps for Android Auto's unique requirements.

\begin{figure}
    \centering
    \includegraphics[width=0.8\linewidth]{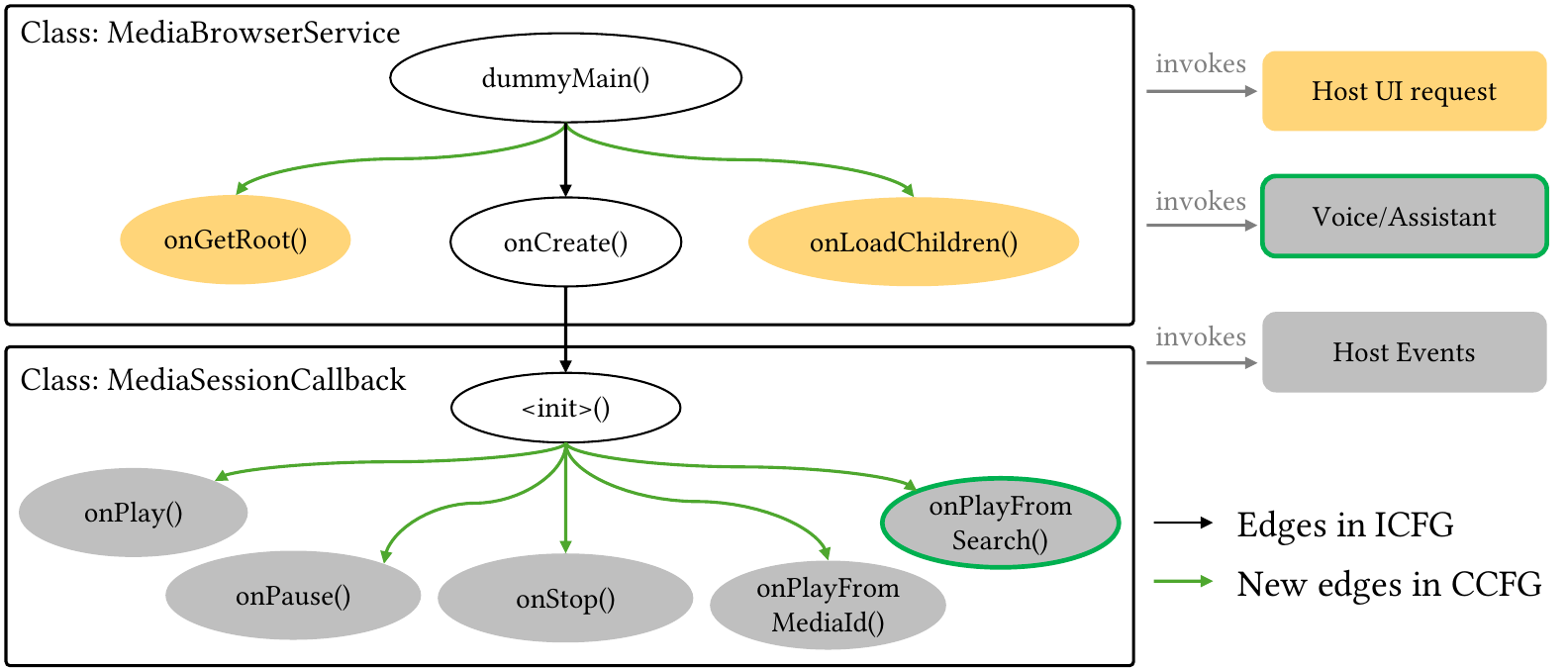}
    \vspace{-10pt}
    \caption{Car Control Flow Graph (CCFG) for Android Auto media apps. 
    Nodes and arrows in black represent standard ICFG code paths; 
    gray nodes indicate required callbacks for UI compliance, and yellow nodes indicate required callbacks for media or voice compliance. 
    Green arrows illustrate host-driven edges added in the CCFG: the Host UI request invokes UI callbacks (\texttt{onGetRoot()}, \texttt{onLoadChildren()}), while Host Events and Voice/Assistant components invoke the required media and voice command callbacks (\texttt{onPlay()}, \texttt{onPause()}, \texttt{onStop()}, \texttt{onPlayFromMediaId()}, and \texttt{onPlayFromSearch()}). }
    \label{fig:ccfg}
    \vspace{-25pt}
\end{figure}

\stitle{State of the Art.} 
Recent work~\cite{auto-messages-testing} has focused on testing message synchronization and UI inconsistencies between an Android phone and a vehicle's screen, but this work primarily addresses messaging functionality and overlooks broader challenges such as UI compliance, media playback, and voice command integration that affect the majority of Android Auto apps.
Current static analysis tools fall short of addressing Android Auto's unique requirements. 
Android Lint~\cite{android-lint} only include a single basic rules for voice command integration but lacks comprehensive coverage of Android Auto-specific compliance checks.
Methodologically, existing static analysis tools such as FlowDroid~\cite{flowdroid} rely on interprocedural control-flow graph (ICFG). 
The ICFG construction misses essential host-driven callbacks, such as \texttt{onGetRoot()} and \texttt{onPlayFromSearch()}. 
The Android Auto runtime invokes these callbacks and are critical for platform compliance. 
This limitation is illustrated in Figure~\ref{fig:ccfg}, which contrasts ICFG with our augmented Car-Control Flow Graph (CCFG). 
The CCFG extends the ICFG by introducing synthetic nodes and edges that explicitly model host-invoked and Assistant-driven flows, manifest and service contracts, and UI template population.
By making these platform-specific interactions visible to static analysis, the CCFG enables comprehensive compliance checking for Android Auto apps and addresses a key gap left by existing tools.

\stitle{Overview of \textsc{AutoComply}.}
In this work, we introduce {\textsc{AutoComply}}, a static analysis tool that detects a broad range of compliance violations unique to Android Auto apps.
At the core {\textsc{AutoComply}} builds on the Car-Control Flow Graph (CCFG), an analysis abstraction that extends conventional control-flow graphs with host-invoked entry points and platform-specific callbacks. 
By capturing interactions across app components, lifecycle methods, and Android Auto APIs, the CCFG enables {\textsc{AutoComply}} to systematically detect compliance violations, including missing or misconfigured callbacks, incomplete voice command support, and UI malfunction, which would otherwise go unnoticed by traditional analysis pipelines.

Building on the CCFG, {\textsc{AutoComply}} implements three dedicated compliance checkers; each checker addresses an important source of platform failures in Android Auto apps found in our study. 
The Media Checker examines the correct handling of playback actions and resource management, focusing on callbacks such as \texttt{onPlay()}, \texttt{onPause()}, and \texttt{onStop()}. 
The UI Checker verifies that required callbacks for UI population (such as \texttt{onGetRoot()} and \texttt{onLoadChildren()}) are correctly implemented, ensuring that media catalogs and navigation menus appear as intended on vehicle displays. 
The Voice Command Checker assesses support for hands-free operation by analyzing Assistant-driven callbacks, most notably \texttt{onPlayFromSearch()}. 
Together, these checkers provide a comprehensive coverage of Android Auto's core compliance requirements, leveraging the precision of the CCFG to highlight missing or faulty implementations.

To evaluate {\textsc{AutoComply}}, we applied it to a diverse set of $31$ Android Auto apps and detected $27$ compliance issues, identifying 13X more issues than Android~Lint~\cite{android-lint} (baseline tool) while no false positives were observed. 
The tool efficiently analyzes apps of varying complexity, demonstrating its applicability. 
Additionally, we reported the detected issues to developers, 14 problems acknowledged, and 8 of them have been fixed by the developers so far. 
By accurately identifying real-world compliance violations, {\textsc{AutoComply}} provides developers with actionable insights to improve platform compliance and the in-vehicle user experience.
The artifact is available at: \url{https://github.com/fakorede/autocomply}.

\stitle{Contributions.}~To summarize, this work makes the following key contributions:
\vspace{-\topsep}
\begin{itemize}[leftmargin=*]
    \item We present the first comprehensive study focused on understanding the challenges faced by developers in creating Android Auto-compliant apps.
    \item We propose Car Control Flow Graph (CCFG), which models the control flow specific to Android Auto apps based on the available features in the apps. 
    \item We developed a novel static analysis specifically designed to identify compliance issues in Android Auto apps, including UI issues and problems with media playback and voice command integration.
    \item We evaluate {\textsc{AutoComply}} on real-world apps, showing its effectiveness by detecting $27$ issues related to Android Auto compliance, $14$ of which were confirmed by developers, and $8$ of them have already been fixed.
\end{itemize}

\section{Formative Study}
\label{sec:study}
To understand the challenges app developers encounter when adapting apps for Android Auto, we conducted a formative study analyzing real-world issues reported in open-source Android Auto app repositories. This study aims to answer:

\vspace{0.5em}
\noindent\textbf{RQ1:} \textit{What are the most prevalent types of Android Auto compliance violations, and what platform-specific factors contribute to their occurrence?}
\vspace{0.5em}

Our findings inform the design of \textsc{AutoComply} by identifying the specific compliance checks needed to detect violations that existing tools miss.

\subsection{Study Corpus Construction}

\textbf{Data Collection.} We collected Android apps from F-Droid~\cite{fdroid}, an open-source Android app repository, resulting in $4,387$ apps. We selected F-Droid because (i) it hosts a large collection of actively maintained open-source Android apps, and (ii) the apps include publicly accessible source code on platforms like GitHub~\cite{github} and GitLab~\cite{gitlab} with issue tracking systems, enabling detailed analysis of reported problems and their fixes.

\stitle{Android Auto App Filtering.} To identify apps with Android Auto support, we analyzed each app's \texttt{AndroidManifest.xml} file for Android Auto-specific declarations. Apps must declare both automotive metadata and \texttt{MediaBrowserService} registration to support Android Auto, as shown in Listing~\ref{lst:auto-manifest}.

\begin{lstlisting}[
    language=XML,
    caption={Android Auto service declaration in manifest file},
    label={lst:auto-manifest},
    numbers=left,
    numberstyle=\tiny,
    basicstyle=\ttfamily\scriptsize,
    backgroundcolor=\color{gray!5!white},
    frame=single,
    framerule=0.5pt,
    rulecolor=\color{gray!40},
    % numbersep=8pt,
    % xleftmargin=2.5em,
    % framexleftmargin=2em,
    captionpos=b
]
<meta-data 
    android:name="com.google.android.gms.car.application"
    android:resource="@xml/automotive_app_desc" />
<service android:name=".MediaBrowserService">
    <intent-filter>
        <action android:name=
          "android.media.browse.MediaBrowserService" />
    </intent-filter>
</service>
\end{lstlisting}

This filtering identified $37$ apps where developers explicitly declared intent to support Android Auto. We refer to this dataset as \textit{\textbf{Corpus-F}}. These apps represent developers' attempts to implement Android Auto support, making them ideal for studying compliance challenges.

\stitle{Issue Collection.} For the $37$ apps in Corpus-F, we mined their GitHub/GitLab repositories to identify Android Auto-specific compliance issues. We performed keyword-based searches on issue titles and descriptions using the following terms: (\texttt{``Android Auto''}, \texttt{``car screen''}, \texttt{``car app''}, \texttt{``automotive''}, \texttt{``MediaBrowserService''}, \texttt{``MediaSession''}, \texttt{``voice command''}, and \texttt{``Google Assistant''}) followed by manual inspection to ensure comprehensive coverage. 


\stitle{Manual Filtering.} Two authors with Android development experience independently reviewed all candidate issues to filter out false positives (e.g., feature requests unrelated to compliance, general Android bugs that coincidentally mention ``Auto''). We retained only issues meeting three criteria: (1) describe failures specific to the Android Auto environment (not reproducible on phones), (2) relate to Android Auto APIs (\texttt{MediaBrowserService}, \texttt{MediaSessionCompat}, voice intents), and (3) involve mandatory platform requirements rather than optional features.

Of the $37$ apps in Corpus-F, we found $98$ Android Auto-specific issues in $14$ apps that had public issue trackers with relevant reports.

\subsection{Analysis Methodology.}

\stitle{Classification Process.} The two authors independently classified each issue based on its manifestation UI (rendering failures), Media (playback control errors), Voice (command handling failures) and root cause following a systematic approach. Initially, each author reviewed all 98 issues and proposed categories based on issue descriptions, code changes in fixes, and affected Android Auto components. The authors then collaboratively refined these categories, merging overlapping ones and clarifying definitions. Next, both authors independently re-classified all issues using the refined definitions.

The issues with classification disagreements were resolved through discussion until consensus was reached.

\stitle{Final Taxonomy.} This systematic process yielded three primary violation categories based on the Android Auto component or the requirement violated:

\begin{itemize}[leftmargin=*]
    \item \textbf{T1 - Media Playback:} Violations related to \texttt{MediaSessionCompat} callbacks and playback control.
    \item \textbf{T2 - User Interface:} Violations related to \texttt{MediaBrowserService} callbacks and content hierarchy.
    \item \textbf{T3 - Voice Commands:} Violations related to the handling of voice action intent and Assistant integration.
\end{itemize}

These categories map directly to Android Auto's documented requirements~\cite{androidauto} for media app compliance, suggesting that our taxonomy captures the platform's core compliance dimensions.

\subsection{Results: Violation Categories}

Our analysis identified 98 Android Auto issues across 14 repositories. Table~\ref{tab:issue-types-causes} presents the complete taxonomy showing issue types, their root causes, and representative examples. Media playback violations dominate (59 issues, 60.2\%), reflecting that most Android Auto apps are media players where playback control is fundamental functionality. UI violations account for 31 issues (31.6\%), while voice command violations represent 8 issues (8.2\%).

\begin{table*}[t]
    \centering
    \caption{Android Auto Issue Types and Root Causes.}
    \label{tab:issue-types-causes}
    \footnotesize
    \vspace{-5pt}
    \renewcommand{\arraystretch}{1.1}
    \setlength{\tabcolsep}{5pt}
    \begin{tabular}{@{}lp{0.20\textwidth}lp{0.47\textwidth}@{}}
        \toprule
        \textbf{Type} & \textbf{Root Cause} & \textbf{Issue Example} & \textbf{Issue Description} \\
        \midrule
        \multirow{3}{*}{\textit{T1: Media Playback}} & Incomplete callback implementation & AntennaPod \#2380 & After completing a podcast episode, the app displays the episode list instead of showing information about the next podcast being played. \\
        \addlinespace[0.3em]
        & State management errors & Vinyl Music Player \#348 & The app crashes when a song is liked using Android Auto. \\

        \midrule
        \multirow{3}{*}{\textit{T2: User Interface}} & Missing callbacks & Harmony-Music \#111 & No user interface displayed in Android Auto. \\
        \addlinespace[0.3em]
        & Incorrect hierarchy structure & Podcini \#57 & The Android Auto UI has been broken for several releases. \\

        \midrule
        \multirow{2}{*}{\textit{T3: Voice Commands}} & Missing intent filter & Ultrasonic \#827 & App fails to support all required voice commands, such as not switching to the requested audio track when a voice command is issued. \\
        \addlinespace[0.3em]
        & Unimplemented callback & Vinyl Music Player \#376 & Required voice commands not supported, such as failing to start the requested track after recognizing the command. \\
        \bottomrule
    \end{tabular}
    \vspace{-1em}
\end{table*}

\subsubsection{T1: Media Playback Violations (60.2\%)}

Media playback violations involve incorrect or incomplete implementation of \texttt{MediaSessionCompat} callbacks that control audio playback in response to user actions in the vehicle interface.

\begin{tcolorbox}[
    colback=blue!3!white,
    colframe=black!60!white,
    boxrule=0.5pt,
    arc=1mm,
    left=3pt,
    right=3pt,
    top=3pt,
    bottom=3pt,
    fontupper=\small
]
\textbf{Platform-Specific Requirement for Media Playback:} Android Auto requires apps to implement MediaSessionCompat.Callback methods (\texttt{onPlay()}, \texttt{onPause()}, \texttt{onStop()}, \texttt{onSkipToNext()}, \texttt{onSkipToPrevious()}, \texttt{onPlayFromMediaId()}) that are invoked by the vehicle system, not by user touches within the app, in response to dashboard button presses or steering wheel controls.
\end{tcolorbox}

\stitle{\textbf{Root Causes.}} As shown in Table~\ref{tab:issue-types-causes}, media playback violations stem from two primary root causes. First, \textit{incomplete callback implementation} occurs when developers implement basic playback callbacks (\texttt{onPlay()}, \texttt{onPause()}) but omit others (\texttt{onSkipToNext()}, \texttt{onSkipToPrevious()}, \texttt{onStop()}). This partial implementation works on phones where users interact with on-screen buttons, but fails in vehicles where all controls must work through callbacks. Second, \textit{state management errors} occur when callbacks fail to properly update playback state (playing, paused, stopped) or metadata (current track, duration), causing the vehicle display to show incorrect information or become unresponsive.

\stitle{\textbf{Why This is Auto-Specific.}} In standard Android apps, playback controls are typically triggered by onClick listeners attached to UI buttons. In Android Auto, the vehicle system invokes callbacks even without the app's UI being visible, requiring a fundamentally different implementation pattern.

\subsubsection{T2: User Interface Violations (31.6\%)}

UI violations involve incorrect or missing implementation of \texttt{MediaBrowserService} callbacks that populate the vehicle's display with browsable content hierarchies.

\begin{tcolorbox}[
    colback=blue!3!white,
    colframe=black!60!white,
    boxrule=0.5pt,
    arc=1mm,
    left=3pt,
    right=3pt,
    top=3pt,
    bottom=3pt,
    fontupper=\small
]
\textbf{Platform-Specific Requirement for User Interface:} Android Auto requires apps to implement \texttt{MediaBrowserService} callbacks (\texttt{onGetRoot()}, \texttt{onLoadChildren()}) that return structured content hierarchies. The vehicle system invokes these callbacks to build the User Interface, apps cannot directly control when or how their content \\ appears.
\end{tcolorbox}

\stitle{\textbf{Root Causes.}} Table~\ref{tab:issue-types-causes} identifies two root causes for UI violations. \textit{Missing callbacks} occur when apps declare MediaBrowserService but fail to implement \texttt{onGetRoot()} or \texttt{onLoadChildren()}, resulting in blank screens. \textit{Incorrect hierarchy structure} happens when apps implement callbacks but return improperly structured MediaItem objects (missing required metadata, incorrect flags, or invalid parent-child relationships), causing rendering failures or navigation errors. 

\stitle{\textbf{Why This is Auto-Specific.}} Standard Android apps control UI rendering through Activities and direct view manipulation \\ (\texttt{findViewById()} and \texttt{setContentView()}). Android Auto inverts this model: apps must describe their content structure through callbacks, and the vehicle system decides when and how to render it. This inverted control model is unique to Android Auto.

\subsubsection{T3: Voice Command Violations (8.2\%)}

Voice command violations involve missing or incorrect handling of voice action intents necessary for hands-free operation.

\begin{tcolorbox}[
    colback=blue!3!white,
    colframe=black!60!white,
    boxrule=0.5pt,
    arc=1mm,
    left=3pt,
    right=3pt,
    top=3pt,
    bottom=3pt,
    fontupper=\small
]
\textbf{Platform-Specific Requirement to avoid Distraction:} Android Auto requires apps to handle \texttt{MEDIA\_PLAY\_FROM\_SEARCH} intents and implement \texttt{onPlayFromSearch()} callback to enable Google Assistant voice commands which is mandatory for safety compliance.
\end{tcolorbox}

\stitle{\textbf{Root Causes.}} As detailed in Table~\ref{tab:issue-types-causes}, voice command violations have two primary causes. \textit{Missing intent filters} occur when apps fail to declare the \texttt{MEDIA\_PLAY\_FROM\_SEARCH} intent filter in their manifest, preventing Google Assistant from routing voice commands to the app. \textit{Unimplemented callbacks} happen when apps declare the intent filter but don't implement \texttt{onPlayFromSearch()}, causing voice commands to be silently ignored.

\stitle{\textbf{Why This is Auto-Specific.}} Voice command integration is optional for standard Android apps but mandatory for Android Auto due to safety requirements: Drivers must be able to control apps without looking at or touching the screen.

\subsection{Discussion and Implications}

\begin{tcolorbox}[colframe=black, arc=0.2mm, boxrule=0.8pt, left=2pt, right=2pt, top=2pt, bottom=2pt]
\textbf{Answer to RQ1:} Our study reveals that Android Auto compliance violations fall into three categories corresponding to the platform's core requirements: media playback control (60.2\%), UI content provision (31.6\%), and voice command handling (8.2\%). These violations mostly stem from Android Auto's inverted control model, where the vehicle system, not the app, initiates callbacks to render UI and handle user interactions.
\end{tcolorbox}

\stitle{\textbf{Comprehensiveness of Categories.}} Our three categories map directly to Android Auto's documented compliance requirements~\cite{androidauto}: T1 corresponds to MediaSession callback requirements, T2 corresponds to MediaBrowser hierarchy requirements, and T3 corresponds to voice action requirements. This mapping suggests our taxonomy captures the platform's fundamental compliance dimensions. While additional edge cases may exist, these three categories represent the mandatory requirements all Android Auto media apps must satisfy.

\stitle{\textbf{Toward Automated Detection.}} Our analysis of 98 Android Auto violations reveals recurring patterns with similar fixes across apps. For example, the missing \texttt{onGetRoot()} and \texttt{onLoadChildren()} UI callbacks (Section~\ref{sec:example}) appears in multiple repositories with identical solutions. This recurrence, combined with evidence of developers referencing fixes from other apps, suggests that compliance knowledge can be generalized and automated. Table~\ref{tab:issue-types-causes} shows that violations cluster around three core requirements with identifiable root causes. We systematically extracted compliance checking rules from these violations and designed \textsc{AutoComply}, targeting Android Auto-specific patterns: missing MediaSession callbacks (T1), incorrect MediaBrowserService implementations (T2), and absent voice command handling (T3).

\section{Motivating Example}
\label{sec:example}

To illustrate the importance of Android Auto compliance checking, we present a real-world case where a subtle implementation gap completely breaks the in-vehicle user experience. 
This example, based on developer reports and issue tracking systems, highlights the type of issue that existing static analysis tools fail to detect, but our {\textsc{AutoComply}} approach can identify.

\stitle{Case Study: Missing Media Browser Interface.}
Consider a music streaming app with implementation shown in Listing~\ref{lst:broken-implementation} that works flawlessly on Android phones but fails to display any user interface when connected to Android Auto (similar to Fig~\ref{fig:ui-issue}). 
Users report that while the app appears in the Android Auto launcher, selecting it results in a blank screen with no visible playlists, albums, or songs. 
This creates a frustrating user experience where drivers are unable to access their music library through the vehicle's infotainment system.

\stitle{The Problem Manifestation.}
When users connect their device to Android Auto and launch the music app, they encounter the following behavior:
\myNum{i}~The app icon is present in the car's infotainment system.
\myNum{ii}~The media browsing interface does not appear.
\myNum{iii}~Background audio playback works if started from the phone, but no in-car controls are available.
\myNum{iv}~Android Auto system logs show connection attempts but do not establish a media hierarchy.

This issue represents a complete failure of Android Auto integration despite the app working fine on the phone itself. 
From a safety perspective, this forces drivers to interact with their phone directly, exactly the behavior Android Auto was designed to prevent.

\begin{figure}
\centering
\begin{minipage}[t]{0.47\textwidth}
\begin{lstlisting}[
    language=Java,
    caption={Problematic implementation},
    label={lst:broken-implementation},
    numbers=left,
    numberstyle=\tiny,
    basicstyle=\ttfamily\scriptsize,
    backgroundcolor=\color{red!5},
    frame=single,
    framerule=0.5pt,
    rulecolor=\color{red!60!black}
]
public class MusicService 
    extends MediaBrowserServiceCompat {
    
    @Override
    public void onCreate() {
        super.onCreate();
        // Initializes MediaSession only
    }
    
    // CRITICAL: Missing onGetRoot() and onLoadChildren()
    // RESULT: Android Auto cannot build UI hierarchy
}
\end{lstlisting}
\end{minipage}
\hfill
\begin{minipage}[t]{0.47\textwidth}
\begin{lstlisting}[
    language=Java,
    caption={Corrected implementation},
    label={lst:fixed-implementation},
    numbers=left,
    numberstyle=\tiny,
    basicstyle=\ttfamily\scriptsize,
    backgroundcolor=\color{green!5},
    frame=single,
    framerule=0.5pt,
    rulecolor=\color{green!60!black}
]
public class MusicService 
    extends MediaBrowserServiceCompat {
    
    @Override
    public void onCreate() {
        super.onCreate();
        // MediaSession initialization
    }

    // NEW: Host-invoked callbacks
    @Override
    public BrowserRoot onGetRoot(...) {
        if (isValidAutoClient(...)) {
            return new BrowserRoot(
                MEDIA_ROOT_ID, null);
        }
        return new BrowserRoot(
            EMPTY_ROOT_ID, null);
    }
    
    @Override
    public void onLoadChildren(...) {
        ...
        result.sendResult(mediaItems);
    }
}
\end{lstlisting}
\end{minipage}
\vspace{-2em}
\caption{Comparison showing the missing callbacks (listing~\ref{lst:broken-implementation}) that cause UI failures and their correct implementation (listing~\ref{lst:fixed-implementation}) that enables proper Android Auto rendering.}
\vspace{-2em}
\label{fig:implementation-comparison}
\end{figure}

\noindent\emph{Root Cause Analysis.}
Our investigation reveals that the application implements a \texttt{MediaBrowserService} but provides incomplete callback implementations. 
Specifically, the developer implemented the service class and basic lifecycle methods but failed to properly implement the media browsing callbacks required for Android Auto's UI rendering.

\noindent\emph{The Correct Implementation.}
Listing~\ref{lst:fixed-implementation} presents the corrected implementation that resolves the user interface issue. 
This version returns valid \texttt{BrowserRoot} objects with the appropriate media hierarchy identifiers and properly implements the asynchronous media loading pattern by always calling \texttt{result.sendResult()}. 
It also includes client validation for security and compliance, and provides a structured media hierarchy that Android Auto can use to generate and display the user interface as intended.

\subsection{Why Existing Tools Cannot Detect This Issue}

Standard static analysis tools fail to detect Android Auto compliance violations due to fundamental architectural mismatches:

\begin{tcolorbox}[
    colback=blue!3!white,
    colframe=black!60!white,
    boxrule=0.5pt,
    arc=1mm,
    left=3pt,
    right=3pt,
    top=3pt,
    bottom=3pt,
    fontupper=\small
]
\textbf{The Core Problem:} Traditional static analyzers like FlowDroid~\cite{flowdroid} model only \textit{app-initiated} control flows, missing the \textit{host-initiated} callback invocations that define Android Auto's execution model.
\end{tcolorbox}

\noindent\textbf{Missing Entry Points.}
FlowDroid~\cite{flowdroid} and similar tools construct ICFGs from standard Android entry points (Activity lifecycle methods). However, critical Android Auto callbacks like \texttt{onGetRoot()} and \texttt{onLoadChildren()} are invoked by the vehicle's infotainment system, not by the app or Android framework. These host-driven callbacks never appear as entry points in traditional analysis.

\noindent\textbf{Incomplete Control Flow.}
Traditional ICFG analysis captures:
\begin{center}
\texttt{onCreate()} $\rightarrow$ \texttt{bindService()} $\rightarrow$ \textit{[analysis stops here]}
\end{center}
The vehicle system's subsequent invocations of \texttt{onGetRoot()} and \texttt{onLoadChildren()} occur outside this flow, making them invisible to analysis.

\noindent\textbf{Superficial Lint Checks.}
Android Lint~\cite{android-lint} includes only basic Auto checks (e.g., voice intent filters, service registration) but doesn't model the service-callback architecture 
required for proper content rendering.

\noindent\textbf{Our Solution.}
{\textsc{AutoComply}} constructs a Car-Control Flow Graph (CCFG) that explicitly models host invocations through synthetic edges (Figure~\ref{fig:ccfg}), capturing the complete execution path including external callback invocations.

\begin{figure*}[t]
    \centering
    \includegraphics[width=0.95\linewidth]{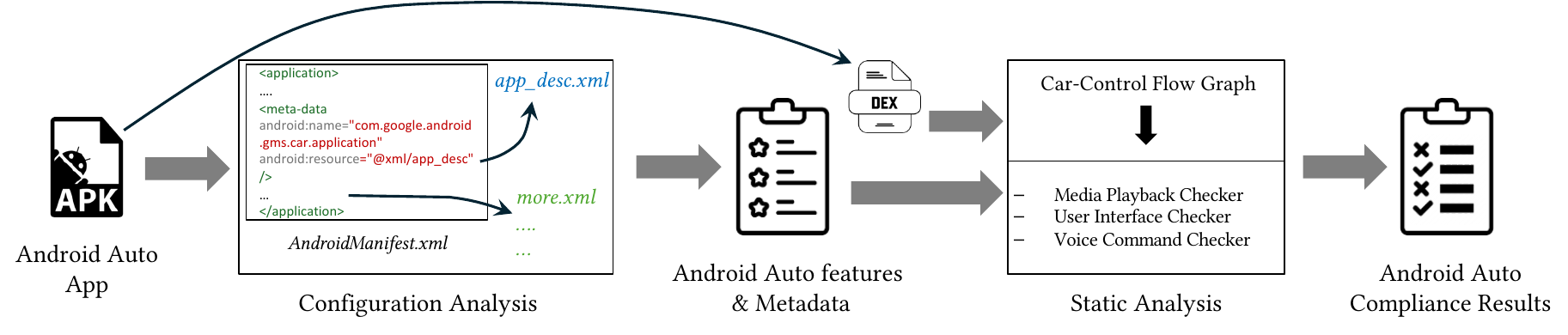}
    \vspace{-10pt}
    \caption{Overview of {\textsc{AutoComply}}: the process starts from configuration analysis to extract Android Auto features and metadata, followed by static analysis, and generating Android Auto compliance results.}
    \vspace{-10pt}
    \label{fig:sys-overview}
\end{figure*}

\subsection{Lessons Learned: Why Android Auto Requires Specialized Analysis}

\noindent\textbf{Challenge 1: Inverted Control Architecture.}
Standard Android apps control UI rendering through direct method calls (\texttt{setContentView()}, \texttt{findViewById()}). Android Auto inverts this model: the vehicle system controls \textit{when} and \textit{how} to request content by invoking app callbacks. Developers must thus ensure their apps respond correctly to these external invocations by implementing auto specific callbacks.

\noindent\textbf{Challenge 2: Safety-Critical Silent Failures.}
Missing callbacks produce blank screens with no error messages, exceptions, or crashes. Drivers cannot access app functionality and must resort to unsafe phone interactions while driving, precisely what Android Auto was designed to prevent.

\noindent\textbf{Challenge 3: Testing Gap.}
These issues pass all conventional validation: compiling without warnings, satisfying Android Lint checks, and working on phones yet fail exclusively in vehicle environments, making them undetectable through standard testing practices.

\noindent\textit{\textbf{Summary.}}
As Figure~\ref{fig:ccfg} illustrates, traditional ICFG approaches fundamentally cannot capture host-driven control flows essential for Android Auto compliance. This motivates our specialized static analysis approach detailed in the following (section~\ref{sec:approach}).

\section{Static Analysis For Android Auto}
\label{sec:approach}
To address the challenges developers face while making their apps Android Auto compliant, we introduce {\textsc{AutoComply}}, a novel static analysis framework designed to systematically detect and analyze issues in Android Auto apps. 
{\textsc{AutoComply}} focuses on the three primary issue types identified in our study: \emph{T1: Media Playback Issues}, \emph{T2: User Interface Issues}, and \emph{T3: Voice Command Issues}. 
For each issue type, {\textsc{AutoComply}} employs dedicated checkers -- \emph{C1, C2, and C3} -- to detect specific compliance violations.
In the following, we describe the architecture of {\textsc{AutoComply}}, the construction of the CCFG, and the compliance checks.

\subsection{Overview of {\textsc{AutoComply}}}
As illustrated in Figure~\ref{fig:sys-overview}, the analysis proceeds in three stages:

\stitle{Configuration Analysis.}
\textsc{AutoComply} begins by parsing configuration artifacts (e.g., \texttt{AndroidManifest.xml}, \texttt{automotive\_app\_desc.xml}) to extract Android Auto-specific metadata, such as \texttt{<meta-data>} entries, service declarations, and intent filters. This step identifies candidate components (e.g., \texttt{MediaBrowserService}, \texttt{MediaSessionCompat}) that indicate support for Android Auto.

\stitle{Car-Control Flow Graph Construction.}
Based on the extracted metadata, the framework extends a traditional ICFG to include \emph{host-driven} control flows. 
These represent interactions where the vehicle's infotainment system or Assistant invokes app callbacks (e.g., \texttt{onGetRoot()}, \texttt{onPlayFromSearch()}) directly. 
The resulting CCFG models both app-initiated and host-initiated execution, enabling whole-app compliance reasoning.

\stitle{Compliance Checking.}
Specialized checkers traverse the CCFG to identify violations of compliance categories identified in our formative study: \emph{(C1)} media playback, \emph{(C2)} User Interface, and \emph{(C3)} voice-command handling. Each checker verifies the presence, connectivity, and semantic correctness of the relevant callbacks.

This analysis pipeline connects configuration data with code-level control flow, producing actionable reports to help developers address compliance issues.

\subsection{Car-Control Flow Graph Construction}
\label{subsec:ccfg}
To enable static reasoning under Android Auto's host-invoked execution model, \emph{Car-Control Flow Graph (CCFG)}  augments the app's control-flow graph with host-driven entry points. 

Given an app $A$ with a traditional ICFG $G_{\text{base}} = (V_{\text{base}}, E_{\text{base}})$, {\textsc{AutoComply}} constructs:
$
  G_{\text{car}} = \left(V_{\text{base}} \cup V_{\text{host}},\; E_{\text{base}} \cup E_{\text{host}}\right),
$

where each node $v_h \in V_{\text{host}}$ represents a callback invoked by the host (e.g., \texttt{onGetRoot()}, \texttt{onLoadChildren()}, \texttt{onPlayFromSearch()}). Each edge $(v_i, v_h) \in E_{\text{host}}$ models a host-triggered invocation, derived by resolving manifest declarations (e.g., \texttt{<meta-data>}, \texttt{<service>}) and mapping them to their implementing service classes.

Figure~\ref{fig:ccfg} illustrates this augmentation. New edges (shown in green) connect system-invoked methods to their corresponding app callbacks, capturing control transfers that traditional static analyses miss. By modeling these host-initiated paths, {\textsc{AutoComply}} enables precise reasoning about execution flows that underlie UI rendering, media playback, and voice-command support.

Algorithm~\ref{algo:carcompat} formalizes this process: Lines 1--9 of Algorithm~\ref{algo:carcompat} capture this construction process. \textsc{ExtractMetadata} reads metadata $autoMetadata$, \textsc{ConstructBaseICFG} builds $G_{base}$, \textsc{InitializeCCFG} prepares the structure, and \textsc{GetCallbacks} enumerates and links Android Auto callbacks to their respective components. 
The following specialized checkers work on lines 10--14 of Algorithm~\ref{algo:carcompat} and check whether these callbacks are implemented according to Android Auto rules, such as returning valid browsing hierarchies and using the correct flags.

\begin{algorithm}[t]
\caption{Android Auto Compliance Analysis}
\label{algo:carcompat}
\SetAlgoLined
\DontPrintSemicolon
\KwIn{An APK file $apk$}
\KwOut{Detected compliance issues}
$autoMetadata \leftarrow \textsc{ExtractMetadata}(apk.manifest)$\;
$G_{base} \leftarrow \textsc{ConstructBaseICFG}(apk)$\;
$ccfg \leftarrow \textsc{InitializeCCFG}(apk)$\;
$requiredCallbacks \leftarrow \textsc{InitializeCallbacks}()$\;
\ForEach{$autoComponent$ \textbf{in} $autoMetadata.components$}{
    $callbacks \leftarrow \textsc{GetCallbacks}(autoComponent)$\;
    $\textsc{AddCallbackNodes}(ccfg, callbacks)$\;
    $\textsc{ConnectCallbackEdges}(ccfg, callbacks)$\;
}
\ForEach{$cb$ \textbf{in} $requiredCallbacks$}{
    \If{$\neg \textsc{Iscompliant}(ccfg, autoComponent, cb)$}{
        report auto compliance issue\;
    }
}

\end{algorithm}

\subsection{Compliance Checkers}
{\textsc{AutoComply}} employs three specialized checkers: \emph{C1} -- Media Playback Checker, \emph{C2} -- User Interface Checker, and \emph{C3} -- Voice Command Checker, to address the specific issue types identified in our study~(Sec. \ref{sec:study}).

\noindent\textit{\textbf{C1 -- Media Playback Checker.}} 
Android Auto requires apps to interact with the vehicle's infotainment system through a specific set of callbacks that are invoked by the platform at runtime. 
These callbacks are not triggered by user actions within the app itself, but rather by the Android Auto runtime in response to driver interactions with the car's interface (screen or hardware buttons) or through voice commands. 
As a result, developers must implement methods such as \texttt{onPlay()}, \texttt{onPause()}, \texttt{onStop()}, \texttt{onSkipToNext()}, \texttt{onSkipToPrevious()}, and \texttt{onPlayFromMediaId()} to handle media playback requests generated externally by the vehicle system. 

To assess compliance, the checker begins by extracting these media-related callbacks as nodes in the CCFG. 
It verifies the presence of all required methods in the \texttt{MediaSessionCompat.Callback} implementation in the app and confirms that each callback is properly connected within the control-flow graph. 
Through dataflow analysis, the checker examines whether these callbacks invoke the appropriate playback or resource management logic, validating that media actions requested by the platform are handled correctly. 
The analysis also evaluates whether the app manages resources such as large media files and buffers effectively within each callback, and whether any error conditions or edge cases are addressed to prevent failures or degraded performance during playback. 
Any missing, incomplete, or incorrectly implemented callbacks, as well as resource management or playback handling issues, are reported as compliance violations. 
All findings are mapped to specific CCFG nodes to provide developers with clear guidance for correcting media playback problems and meeting Android Auto compliance requirements.

\noindent\textit{\textbf{C2 -- User Interface Checker.}} 
In Android Auto, vehicle's infotainment relies on callbacks such as \texttt{onGetRoot()} and \texttt{onLoadChildren()} to dynamically populate and display media or navigation hierarchies for the driver. 
The checker begins by identifying the relevant \texttt{MediaBrowserServiceCompat} components in the app and locating the corresponding nodes for \texttt{onGetRoot()} and \texttt{onLoadChildren()} in the CCFG. 
It then checks that both methods are present with the correct signatures and analyzes their control-flow to verify that all execution paths return valid \texttt{BrowserRoot} objects (in \texttt{onGetRoot()}) and that \texttt{sendResult()} is called as required in \texttt{onLoadChildren()}. 

The checker also validates that the app provides a non-empty and properly structured hierarchy of media or navigation items, which is essential for Android Auto to generate a usable user interface on the dashboard. 
In addition, the analysis reviews client validation logic and error handling to identify cases that could result in blank screens, empty lists, or unresponsive menus for the driver. 
Any deficiencies, such as missing callbacks, incorrect method signatures, incomplete media hierarchies, or weak error handling, are flagged as user interface compliance violations.

\noindent\textit{\textbf{C3: Voice Command Checker.}} 
For safety and distraction-free driving, Android Auto requires apps to fully support voice interactions so that drivers can control playback and navigation without taking their hands off the wheel. 
The Voice Command Compliance Checker uses the CCFG to evaluate whether an app implements the platform's requirements for voice command support. 
In the Android Auto environment, the system invokes \texttt{onPlayFromSearch()} callback, in response to driver voice commands or requests through the Google Assistant. 
Unlike callbacks in standard Android apps, these methods are triggered externally by the platform rather than through direct user interaction within the app interface. 

To evaluate compliance, the checker identifies all Assistant-driven callbacks in the CCFG, focusing primarily on \texttt{onPlayFromSearch()} within the app's \texttt{MediaSessionCompat.Callback} implementation. 
It verifies the presence and correct signature of these methods, and applies control-flow analysis to ensure that voice command inputs are processed correctly. 
The checker also examines whether command parameters are validated, playback state is updated appropriately, and errors or unrecognized requests are handled gracefully to avoid silent failures. 
Any missing, incomplete, or incorrectly implemented voice command handlers, as well as improper handling of command inputs or error conditions, are reported as voice compliance violations.

\section{Evaluation}
\label{sec:eval}
This section evaluates {\textsc{AutoComply}} on real-world Android Auto apps, demonstrating its effectiveness in detecting compliance violations, its advantages over existing tools, and its practical value to developers. We address the following research questions:

\begin{itemize}[leftmargin=*]
    \item \textbf{RQ2 (Effectiveness):} Can {\textsc{AutoComply}} effectively and precisely detect Android Auto compliance issues in real-world apps?
    \item \textbf{RQ3 (Comparison):} How does {\textsc{AutoComply}} compare with existing tools in coverage and accuracy?
    \item \textbf{RQ4 (Usefulness):} Do developers acknowledge and address issues detected by {\textsc{AutoComply}}?
\end{itemize}

\begin{table*}[t]
    \centering
    \caption{Detailed static analysis results when Android Lint and {\textsc{AutoComply}} applied on selected Android Auto apps. Stars: Number of GitHub stars; Downloads: Downloads on Google Play; KLOC: Source code size (thousands of lines); Commit: Latest commit hash; TP: True Positives; FP: False Positives; Time(s): Analysis time in seconds.}
    \vspace{-5pt}
    \label{tab:eval-effectiveness}
    \small
    \begin{tabular}{@{}rlrrrcrrrrrr@{}}
        \toprule
        & & & & & & \multicolumn{3}{c}{\textbf{Android Lint}} & \multicolumn{3}{c}{\textbf{{\textsc{AutoComply}}}} \\ 
        \cmidrule(lr){7-9} \cmidrule(lr){10-12}
        \textbf{\#} & \textbf{Repository} & \textbf{Stars} & \textbf{Downloads} & \textbf{KLOC} & \textbf{Commit} & \textbf{TP} & \textbf{FP} & \textbf{Time} & \textbf{TP} & \textbf{FP} & \textbf{Time} \\ 
        \midrule
1 & anandnet/Harmony-Music~\cite{harmony-music} & 1.4K & -- & 35 & ff91d19 & 0 & 0 & 1 & 3 & 0 & 1 \\
2 & antoinepirlot/Satunes~\cite{satunes} & 38 & -- & 138.8 & 24e3933 & 0 & 0 & 1 & 2 & 0 & 18 \\
3 & DJDoubleD/refreezer~\cite{refreezer} & 433 & 49K & 34.7 & aecf242 & 0 & 0 & 1 & 1 & 0 & 4 \\
4 & gokadzev/Musify~\cite{musify} & 2.4K & 467K & 9.5 & f524674 & 0 & 0 & 1 & 3 & 0 & 2 \\
5 & jellyfin/jellyfin-android~\cite{jellyfin-android} & 1.7K & 1M+ & 158.9 & 7886df9 & 0 & 0 & 140 & 2 & 0 & 7 \\
6 & KRTirtho/spotube~\cite{spotube} & 33.9K & -- & 68.2 & 8c1337d & 0 & 0 & 1 & 3 & 0 & 2 \\
7 & LISTEN-moe/android-app~\cite{listen-moe} & 258 & -- & 90.4 & 4194aeb & 0 & 0 & 30 & 2 & 0 & 3 \\
8 & namidaco/namida~\cite{namida} & 2.9K & 289K & 92.9 & 675dc79 & 0 & 0 & 1 & 2 & 0 & 3 \\
9 & nextcloud/news-android~\cite{news-android} & 705 & 5K+ & 223.5 & 9636d61 & 0 & 0 & 18 & 1 & 0 & 5 \\
10 & nt4f04uNd/sweyer~\cite{sweyer} & 209 & -- & 27.5 & e5abe5a & 0 & 0 & 1 & 3 & 0 & 2 \\
11 & OxygenCobalt/Auxio~\cite{auxio} & 2.3K & -- & 47.9 & fce77ec & 1 & 0 & 198 & 1 & 0 & 6 \\
12 & quran/quran\_android~\cite{quran-android} & 2.1K & 50M+ & 58.2 & f5bd3dc & 0 & 0 & 138 & 1 & 0 & 6 \\
13 & Simple-Music-Player~\cite{simplemusicplayer} & 1.3K & -- & 215.3 & 498086a & 0 & 0 & 64 & 1 & 0 & 8 \\
14 & sosauce/CuteMusic~\cite{cutemusic} & 363 & 4K & 30.4 & 97bb20b & 1 & 0 & 119 & 1 & 0 & 2 \\
15 & timusus/Shuttle2~\cite{shuttle2} & 208 & -- & 295.5 & 6fd520f & 0 & 0 & 315 & 1 & 0 & 21 \\ 
        \midrule
\multicolumn{6}{@{}l}{\textbf{Total Issues / Geomean Time (s)}} & \textbf{2} & \textbf{0} & \textbf{11.1} & \textbf{27} & \textbf{0} & \textbf{4.2} \\
        \bottomrule
    \end{tabular}
    \vspace{-1em}
\end{table*}

\subsection{Methodology}
To evaluate compliance violations in Android Auto apps, we compiled a dataset of open-source apps from GitHub~\cite{github} by performing a code search for repositories with Android Auto related declarations in their \texttt{AndroidManifest.xml}. The keywords we used are \texttt{com.google.android.gms.car.application} and \\ \texttt{android.media.browse.MediaBrowserService}.
We collected apps that \myNum{i} have commits after January 2024 (i.e., actively-maintained), \myNum{ii} have at least 30 stars and are not forks, \myNum{iii} have a public issue tracking system (for submitting issue reports). Finally, we manually filtered the results for duplicates, library repositories, and plugins.

This dataset, \emph{Corpus-G}, consists of $31$ Android Auto apps with a total of $6.5$ million lines of code.
The availability of source code enables a direct comparison between {\textsc{AutoComply}} and Android Lint~\cite{android-lint}, facilitating a comprehensive assessment of their detection capabilities.
The collected apps differ in functionality, size, and complexity, providing a well-rounded sample for analyzing compliance challenges in Android Auto development.

Android Lint~\cite{android-lint} is a static analysis tool provided by Google to detect potential issues in Android apps. 
It includes basic compliance checks for Android Auto, primarily focusing on media playback compliance. 
In our evaluation, we use Android Lint as a baseline to compare its issue detection capabilities against {\textsc{AutoComply}}, highlighting differences in coverage and accuracy.

Our experiments were conducted on an Apple MacBook Pro equipped with Intel Core i7 2.7 GHz Quad-Core processors and 16GB of memory, reflecting a typical system configuration accessible to most app developers.
For testing on a physical device, we utilized a Google Pixel 7a running Android 14.0.

\subsection{RQ2: Effectiveness of {\textsc{AutoComply}}}

To evaluate the applicability of AutoComply, we applied it to
Corpus-G, a dataset of $31$ open-source Android Auto apps spanning diverse categories. The objective was to assess whether AutoComply could effectively detect compliance issues across various
real-world apps and remain computationally efficient for large scale analysis. Our analysis identified 27 compliance issues across 15 apps, spanning media playback, UI constraints, and voice command handling. These findings demonstrate that Android Auto compliance challenges are common even in actively maintained apps. The geometric
mean analysis time per app was 4.2 seconds, showing that AutoComply can efficiently analyze apps of varying complexity without significant overhead.

The evaluation followed a two-step process. First, we performed static analysis on the selected apps to identify potential compliance violations. Next, we tested these apps in a real Vehicle to verify the detected issues in a real execution environment.

All identified issues were manually validated, confirming their impact on app behavior. 
As discussed later in this paper, the verified issues were also reported to developers, reinforcing the practical relevance of {\textsc{AutoComply}}. 
This evaluation examines the tool's ability to detect Auto compliance violations accurately while minimizing incorrect detections. 
The results are summarized in Table~\ref{tab:eval-effectiveness}, under the \textsc{AutoComply} columns.

\stitle{Issue Detection Accuracy.}
Across the evaluated apps, {\textsc{AutoComply}} correctly identified 27 compliance issues without reporting any false positives or missing any expected violations. 
The absence of false negatives shows that the tool successfully captures all relevant compliance issues, while the lack of false positives indicates that its detections are highly reliable. 
These results highlight {\textsc{AutoComply}}'s strong detection capabilities, providing developers with precise and actionable findings.

\stitle{Discussion.}
As mentioned previously, cross-platform apps may implement Android Auto functionality differently, requiring additional support in {\textsc{AutoComply}} to handle such variations. Similarly, apps incorporating C/C++ may introduce compliance-related logic outside the primary Java/Kotlin codebase, necessitating further enhancements.
While {\textsc{AutoComply}} performs exceptionally well for Java/Kotlin-based Android Auto apps, its applicability to multi-language projects requires further evaluation. 
Many modern Android Auto apps integrate JavaScript (e.g., React-Native), Dart (i.e. Flutter) for functionalities such as UI rendering or media playback. 
Since {\textsc{AutoComply}} primarily analyzes Java/Kotlin code, it may not detect compliance-related logic implemented in other languages, potentially leading to false positives in hybrid apps.

\begin{tcolorbox}[colframe=black, arc=0.2mm, boxrule=0.8pt, left=2pt, right=2pt, top=2pt, bottom=2pt]
\textbf{Answer to RQ2:} {\textsc{AutoComply}} achieves high accuracy in detecting Android Auto compliance issues, correctly identifying 27 issues with no false positives or false negatives. 

These results demonstrate that {\textsc{AutoComply}} provides reliable checks on real-world Android Auto apps.
\end{tcolorbox}

\subsection{RQ3: Comparison with Existing Tools}
To evaluate {\textsc{AutoComply}} against existing Android Auto analysis tools, we compare it with Android Lint~\cite{android-lint}, which, to the best of our knowledge, is the only tool that includes some built-in checks for Android Auto compliance violations, specifically for verifying that the \texttt{MediaBrowserService} has the necessary intent filters declared in the app's manifest. 
Due to its native support for Android Auto-related analysis, we use Android Lint as the baseline for comparison.

\stitle{Issue Coverage and Accuracy.}
The results in Table~\ref{tab:eval-effectiveness} demonstrate that {\textsc{AutoComply}} provides significantly broader issue coverage than Android Lint. 
While Android Lint detected only $2$ compliance issues, {\textsc{AutoComply}}~identified $27$ issues across the same set of 15 apps - a 13X increase. 
This suggests that Android Lint's built-in compliance checks, which focus only on voice search issues, may overlook many real-world problems. 
In contrast, {\textsc{AutoComply}} is specifically designed to capture Android Auto-related constraints, including UI constraints, media playback and voice command handling, which are critical for compliance. In addition, {\textsc{AutoComply}} was able to detect both issues detected by Android Lint.

Both tools reported zero false positives, meaning that every detected issue was a valid compliance violation. However, \textsc{AutoComply}'s ability to detect a broader range of issues demonstrates its higher sensitivity to real compliance problems that Android Lint does not identify.

\stitle{Performance Efficiency.}
{\textsc{AutoComply}} achieves significantly better performance, analyzing apps in a geometric mean of $4.2$ seconds, compared to $11.1$ seconds for Android Lint- a more than 2X speed improvement. 
This efficiency makes {\textsc{AutoComply}} more practical for real-world use, allowing developers to detect issues quickly with minimal overhead.

This performance gain is due to fundamental differences in how the tools operate. 
Android Lint analyzes source code and requires processing to multiple source files, which increases processing time. 
In contrast, {\textsc{AutoComply}} operates directly on APK files and leverages the CCFG to model app execution more efficiently. 
This enables faster and more precise detection of compliance issues, making {\textsc{AutoComply}} both more effective and computationally efficient.

\begin{tcolorbox}[colframe=black, arc=0.2mm, boxrule=0.8pt, left=2pt, right=2pt, top=2pt, bottom=2pt]
\textbf{Answer to RQ3:} \textsc{\textsc{AutoComply}} significantly outperforms Android Lint by detecting 27 issues compared to only 2, while both tools report zero false positives. 
Additionally, {\textsc{AutoComply}} runs more than 2X faster, making it a more comprehensive and efficient solution for Android Auto compliance analysis.
\end{tcolorbox}

\subsection{RQ4: Usefulness of {\textsc{AutoComply}}}

To evaluate whether developers find {\textsc{AutoComply}}'s detected violations actionable, we reported all 27 compliance issues to their respective app developers through their GitHub issue trackers. Each report followed a structured format including: (1) clear issue description with observable symptoms, (2) root cause analysis identifying the missing or misconfigured Android Auto API, (3) links to official Android Auto documentation, and (4) references to successful implementations in similar apps. To avoid overwhelming maintainers, we consolidated all detected issues for each app into a single, well-documented bug report.

\stitle{Discovered Issues Overview.} {\textsc{AutoComply}} detected 27 compliance violations across 15 apps, distributed as follows: 10 UI violations (T2), 7 media playback violations (T1), and 10 voice command violations (T3). The detected issues span all three violation categories from our taxonomy, validating that {\textsc{AutoComply}} provides comprehensive coverage of Android Auto compliance requirements. UI violations primarily involved missing \texttt{onGetRoot()} or \texttt{onLoadChildren()} callbacks that prevented content from displaying on vehicle screens. Media playback violations included incomplete callback implementations (e.g., missing \texttt{onStop()}) that caused unexpected behavior during playback control. Voice command violations consisted of missing intent filters or unimplemented \texttt{onPlayFromSearch()} callbacks that prevented Google Assistant integration. Notably, several apps had multiple violation types, for instance, some apps had both UI rendering issues and voice command gaps, indicating that Android Auto compliance requires attention across multiple platform requirements simultaneously. All detected violations matched patterns identified in our formative study, demonstrating that {\textsc{AutoComply}} successfully generalizes compliance checking rules beyond the training dataset.

\stitle{Developer Response and Engagement.} Of the 27 reported issues, developers have responded to 14 so far. Eight issues have been fixed and merged into production, while six acknowledged issues are scheduled for future releases.

Developer responses fell into three patterns. First, some developers responded enthusiastically with immediate fixes, releasing updated versions and requesting verification that issues were resolved. These developers treated the reports as high-priority user experience improvements. Second, other developers acknowledged the issues positively but indicated resource constraints, expressing appreciation for the detailed reports with documentation references and noting plans to address the issues when time permits. One developer described Android Auto support as requiring 'minimal tinkering' and committed to keeping it on the development radar. Third, some developers sought community assistance due to time constraints, with one requesting pull request contributions, a response pattern enabled by our detailed reports with working examples.

Notably, one developer acknowledged the potential user experience improvement but expressed broader concerns about framework dependencies that influenced their development priorities. Despite varying circumstances, all fixed issues were addressed within one to four weeks, demonstrating that {\textsc{AutoComply}} identifies actionable problems developers find valuable. The response variation reflects typical open-source dynamics where maintainer availability and project priorities influence fix timelines rather than issue validity.

\stitle{Impact on App Quality.} For the 8 fixed issues, we verified that fixes were correctly implemented and resolved the compliance violations. We re-ran {\textsc{AutoComply}} on updated versions and confirmed zero violations in the previously flagged categories. These fixes directly improved user experience by enabling previously broken functionality, apps that showed blank screens now properly display content hierarchies, media playback controls now function correctly across all callbacks, and voice commands now work as intended. The fixes demonstrate that {\textsc{AutoComply}} detects genuine compliance violations that, once addressed, measurably improve Android Auto app quality and safety.

\begin{tcolorbox}[colframe=black, arc=0.2mm, boxrule=0.8pt, left=2pt, right=2pt, top=2pt, bottom=2pt]
\textbf{Answer to RQ4:} {\textsc{AutoComply}} effectively identifies actionable compliance violations that developers find valuable. All fixed issues were verified to resolve the detected violations. The positive reception, rapid fix adoption, and measurable quality improvements demonstrate that {\textsc{AutoComply}} provides practical value to developers.
\end{tcolorbox}

\section{Threats to Validity}
\label{sec:threats}

\stitle{External Validity.}  
\textsc{AutoComply} was primarily evaluated on apps written in Java and Kotlin, limiting its applicability to cross-platform apps using C\#, Xamarin, or native C/C++. While many of these platforms do not currently support Android Auto, their exclusion may affect the generalizability of our results.

\stitle{Construct Validity.}  
Our evaluation focuses on specific Android Auto compliance issues but does not cover runtime performance or user interaction patterns, meaning certain issues may go undetected. Code obfuscation also poses a challenge, as it modifies class names and control flows, potentially impacting {\textsc{AutoComply}}' accuracy and leading to false positives.

\stitle{Conclusion Validity.}  
Analysis time varies with app complexity, with more intricate apps requiring additional processing time. While this does not affect detection accuracy, longer runtimes could impact developer workflows. Additionally, although we include widely used apps in our dataset, the corpus may not fully represent the entire Android Auto ecosystem, which could influence the broader applicability of our conclusions.

\section{Related Work}
\label{sec:related}
Mobile app reliability and compatibility issues have been extensively studied in the Android ecosystem~\cite{cai2019large, huang2018understanding, li2018cid, liu2021identifying, scalabrino2019data, wang2019characterizing, wei2016taming}. Wei et al.~\cite{wei2016taming} characterized fragmentation-induced compatibility issues and proposed FicFinder for static detection. Their follow-up work, Pivot~\cite{wei2018understanding}, identifies compatibility problems across Android versions and devices. Huang et al.~\cite{huang2018understanding} developed CIDER to detect callback compatibility issues from API evolution, while Li et al.~\cite{li2018cid} created CiD to identify incompatibilities from API changes. These works focus on device fragmentation and OS version compatibility within standard Android, whereas our work addresses platform-specific compliance for Android Auto's unique service-based architecture where the vehicle system invokes app callbacks externally.

Android Auto-specific research remains limited. Zhang et al.~\cite{auto-messages-testing} developed techniques to test message synchronization and UI consistency between phones and vehicle dashboards, focusing primarily on messaging functionality. While valuable, their work does not address broader compliance challenges including service-based architecture requirements, media playback callbacks, or voice command integration that affect most Android Auto apps. Mandal et al.~\cite{mandal2019static, mandal2018vulnerability} explored security vulnerabilities in Android Auto apps, focusing on privacy leaks and malicious behaviors. Their work assumes functional compliance and targets security issues, whereas we address the fundamental implementation gaps preventing apps from working correctly in Android Auto environments.

Static analysis tools like FlowDroid~\cite{flowdroid} provide precise taint analysis for Android apps through ICFG construction. However, as we demonstrate in Section~\ref{sec:example}, traditional ICFGs fundamentally cannot capture host-driven callbacks essential for Android Auto compliance because they only model app-initiated flows. Android Lint~\cite{android-lint} includes basic Android Auto checks, primarily verifying manifest declarations, but our evaluation (Section~\ref{sec:eval}) shows it detects only $2$ issues compared to 
{\textsc{AutoComply}}'s $27$. Lint's rule-based approach lacks the deep control-flow analysis 
necessary to detect missing callback implementations or media playback handlers.

Other research has addressed race conditions and energy bugs~\cite{hsiao2014race, maiya2014race,hu2011automating}, network and GPS issues~\cite{liang2014caiipa}, app quality~\cite{liu2014characterizing,lillack2014tracking,muccini2012software,berardinelli2010performance,arijo2011modular}, configuration changes~\cite{runtimedroid,livedroid,ast2023}, and memory leaks~\cite{yan2013systematic}. While these contributions improve mobile app quality, they target general Android problems rather than platform-specific compliance requirements unique to Android Auto.

Our work provides the first comprehensive study of Android Auto compliance through analysis of $98$ real-world issues, introduces the Car-Control Flow Graph (CCFG) to model host-invoked callbacks invisible to traditional analysis, and demonstrates practical impact with $27$ detected violations and $8$ developer-confirmed fixes. Unlike existing tools that focus on general Android compatibility or security, 
{\textsc{AutoComply}} systematically detects compliance violations specific to Android Auto's inverted control architecture.

\section{Conclusion}
\label{sec:conclusion}
Despite the vast number of Android apps available, numbering in the millions, the number of Android Auto-compliant apps remains remarkably small.
This disparity highlights the significant challenges that developers face in adapting their apps for the automotive platform. 
In this paper, we conducted a comprehensive investigation into these challenges.
Our study revealed that UI issues, media playback errors, and voice command integration errors are often caused by incompatibility, and developers are left to carry out manual issue detection. 
To address these challenges, we introduced {\textsc{AutoComply}}, a novel static analysis designed to detect compliance issues in Android Auto apps. 
{\textsc{AutoComply}} applies static analysis techniques to identify potential problems, providing developers with actionable insights to enhance the reliability of their apps. 
Our evaluation of {\textsc{AutoComply}} on 31 Android apps demonstrated its effectiveness, identifying 27 compliance issues.
This work contributes to the limited body of research on Android Auto development by providing a practical solution for detecting compliance issues early in the development process. 
Moving forward, we aim to extend the capabilities of {\textsc{AutoComply}} to cover additional features and add support for dynamic analysis to detect runtime issues.

\section*{Acknowledgment}
This material is based upon the work supported in part by the National Science Foundation (NSF) under Grant No.~2449694 and the Louisiana Board of Regents. Any opinions, findings, and conclusions or recommendations expressed in this material are those of the authors and do not necessarily reflect the views of the NSF or the Louisiana Board of Regents.

\balance
\bibliographystyle{ACM-Reference-Format}
\bibliography{sample-base}


\begin{thebibliography}{47}


\ifx \showCODEN    \undefined \def \showCODEN     #1{\unskip}     \fi
\ifx \showDOI      \undefined \def \showDOI       #1{#1}\fi
\ifx \showISBNx    \undefined \def \showISBNx     #1{\unskip}     \fi
\ifx \showISBNxiii \undefined \def \showISBNxiii  #1{\unskip}     \fi
\ifx \showISSN     \undefined \def \showISSN      #1{\unskip}     \fi
\ifx \showLCCN     \undefined \def \showLCCN      #1{\unskip}     \fi
\ifx \shownote     \undefined \def \shownote      #1{#1}          \fi
\ifx \showarticletitle \undefined \def \showarticletitle #1{#1}   \fi
\ifx \showURL      \undefined \def \showURL       {\relax}        \fi
\providecommand\bibfield[2]{#2}
\providecommand\bibinfo[2]{#2}
\providecommand\natexlab[1]{#1}
\providecommand\showeprint[2][]{arXiv:#2}

\bibitem[anandnet(2023)]%
        {harmony-music}
\bibfield{author}{\bibinfo{person}{anandnet}.} \bibinfo{year}{2023}\natexlab{}.
\newblock \bibinfo{title}{A cross platform app for streaming music.}
\newblock
  \bibinfo{howpublished}{\url{https://github.com/anandnet/Harmony-Music/}}.
\newblock
\newblock
\shownote{Accessed: 2024-08-14}.


\bibitem[Antoinepirlot(2024)]%
        {satunes}
\bibfield{author}{\bibinfo{person}{Antoinepirlot}.}
  \bibinfo{year}{2024}\natexlab{}.
\newblock \bibinfo{title}{Modern MP3 Player to listen to your local music files
  on Android Lollipop 5.1.1+ \& compatible with Android Auto.}
\newblock
  \bibinfo{howpublished}{\url{https://github.com/antoinepirlot/Satunes}}.
\newblock
\newblock
\shownote{Accessed: 2024-08-14}.


\bibitem[Arijo et~al\mbox{.}(2011)]%
        {arijo2011modular}
\bibfield{author}{\bibinfo{person}{Niaz Arijo}, \bibinfo{person}{Reiko Heckel},
  \bibinfo{person}{Mirco Tribastone}, {and} \bibinfo{person}{Stephen Gilmore}.}
  \bibinfo{year}{2011}\natexlab{}.
\newblock \showarticletitle{Modular performance modelling for mobile
  applications}. In \bibinfo{booktitle}{\emph{ACM SIGSOFT Software Engineering
  Notes}}, Vol.~\bibinfo{volume}{36}. ACM, \bibinfo{pages}{329--334}.
\newblock


\bibitem[Arzt et~al\mbox{.}(2014)]%
        {flowdroid}
\bibfield{author}{\bibinfo{person}{Steven Arzt}, \bibinfo{person}{Siegfried
  Rasthofer}, \bibinfo{person}{Christian Fritz}, \bibinfo{person}{Eric Bodden},
  \bibinfo{person}{Alexandre Bartel}, \bibinfo{person}{Jacques Klein},
  \bibinfo{person}{Yves Le~Traon}, \bibinfo{person}{Damien Octeau}, {and}
  \bibinfo{person}{Patrick McDaniel}.} \bibinfo{year}{2014}\natexlab{}.
\newblock \showarticletitle{FlowDroid: Precise Context, Flow, Field,
  Object-sensitive and Lifecycle-aware Taint Analysis for Android Apps}. In
  \bibinfo{booktitle}{\emph{Proceedings of the 35th ACM SIGPLAN Conference on
  Programming Language Design and Implementation}} (Edinburgh, United Kingdom)
  \emph{(\bibinfo{series}{PLDI '14})}. \bibinfo{publisher}{ACM},
  \bibinfo{address}{New York, NY, USA}, \bibinfo{pages}{259--269}.
\newblock
\showISBNx{978-1-4503-2784-8}
\urldef\tempurl%
\url{https://doi.org/10.1145/2594291.2594299}
\showDOI{\tempurl}


\bibitem[Berardinelli et~al\mbox{.}(2010)]%
        {berardinelli2010performance}
\bibfield{author}{\bibinfo{person}{Luca Berardinelli},
  \bibinfo{person}{Vittorio Cortellessa}, {and} \bibinfo{person}{Antinisca
  Di~Marco}.} \bibinfo{year}{2010}\natexlab{}.
\newblock \showarticletitle{Performance modeling and analysis of context-aware
  mobile software systems}.
\newblock \bibinfo{journal}{\emph{Fundamental Approaches to Software
  Engineering}} (\bibinfo{year}{2010}), \bibinfo{pages}{353--367}.
\newblock


\bibitem[Cai et~al\mbox{.}(2019)]%
        {cai2019large}
\bibfield{author}{\bibinfo{person}{Haipeng Cai}, \bibinfo{person}{Ziyi Zhang},
  \bibinfo{person}{Li Li}, {and} \bibinfo{person}{Xiaoqin Fu}.}
  \bibinfo{year}{2019}\natexlab{}.
\newblock \showarticletitle{A large-scale study of application
  incompatibilities in android}. In \bibinfo{booktitle}{\emph{Proceedings of
  the 28th ACM SIGSOFT International Symposium on Software Testing and
  Analysis}}. \bibinfo{pages}{216--227}.
\newblock


\bibitem[Ceci(2023)]%
        {number-of-apps-on-stores}
\bibfield{author}{\bibinfo{person}{L. Ceci}.} \bibinfo{year}{2023}\natexlab{}.
\newblock \bibinfo{title}{Number of apps available in leading app store}.
\newblock
  \bibinfo{howpublished}{\url{https://www.statista.com/statistics/276623/number-of-apps-available-in-leading-app-stores/}}.
\newblock
\newblock
\shownote{Accessed: 2023-11-06}.


\bibitem[Developers(2015)]%
        {androidauto}
\bibfield{author}{\bibinfo{person}{Android Developers}.}
  \bibinfo{year}{2015}\natexlab{}.
\newblock \bibinfo{title}{Android Auto}.
\newblock \bibinfo{howpublished}{\url{https://www.android.com/auto/}}.
\newblock
\newblock
\shownote{Accessed: 2024-09-03}.


\bibitem[DJDoubleD({[n.\,d.]})]%
        {refreezer}
\bibfield{author}{\bibinfo{person}{DJDoubleD}.}
  \bibinfo{year}{[n.\,d.]}\natexlab{}.
\newblock \bibinfo{title}{An alternative Deezer music streaming \& downloading
  client, based on Freezer.}
\newblock \bibinfo{howpublished}{\url{https://github.com/DJDoubleD/refreezer}}.
\newblock
\newblock
\shownote{Accessed: 2024-08-14}.


\bibitem[F-Droid(2010)]%
        {fdroid}
\bibfield{author}{\bibinfo{person}{F-Droid}.} \bibinfo{year}{2010}\natexlab{}.
\newblock \bibinfo{title}{F-Droid}.
\newblock \bibinfo{howpublished}{\url{https://f-droid.org/en/packages/}}.
\newblock
\newblock
\shownote{Accessed: 2024}.


\bibitem[Farooq and Zhao(2018)]%
        {runtimedroid}
\bibfield{author}{\bibinfo{person}{Umar Farooq} {and} \bibinfo{person}{Zhijia
  Zhao}.} \bibinfo{year}{2018}\natexlab{}.
\newblock \showarticletitle{RuntimeDroid: Restarting-Free Runtime Change
  Handling for Android Apps}. In \bibinfo{booktitle}{\emph{Proceedings of the
  16th Annual International Conference on Mobile Systems, Applications, and
  Services}} (Munich, Germany) \emph{(\bibinfo{series}{MobiSys '18})}.
  \bibinfo{publisher}{Association for Computing Machinery},
  \bibinfo{address}{New York, NY, USA}, \bibinfo{pages}{110–122}.
\newblock
\showISBNx{9781450357203}
\urldef\tempurl%
\url{https://doi.org/10.1145/3210240.3210327}
\showDOI{\tempurl}


\bibitem[Farooq et~al\mbox{.}(2020)]%
        {livedroid}
\bibfield{author}{\bibinfo{person}{Umar Farooq}, \bibinfo{person}{Zhijia Zhao},
  \bibinfo{person}{Manu Sridharan}, {and} \bibinfo{person}{Iulian Neamtiu}.}
  \bibinfo{year}{2020}\natexlab{}.
\newblock \showarticletitle{LiveDroid: Identifying and Preserving Mobile App
  State in Volatile Runtime Environments}.
\newblock \bibinfo{journal}{\emph{Proc. ACM Program. Lang.}}
  \bibinfo{volume}{4}, \bibinfo{number}{OOPSLA}, Article
  \bibinfo{articleno}{160} (\bibinfo{date}{nov} \bibinfo{year}{2020}),
  \bibinfo{numpages}{30}~pages.
\newblock
\urldef\tempurl%
\url{https://doi.org/10.1145/3428228}
\showDOI{\tempurl}


\bibitem[Github(2016)]%
        {github}
\bibfield{author}{\bibinfo{person}{Github}.} \bibinfo{year}{2016}\natexlab{}.
\newblock \bibinfo{howpublished}{\url{https://github.com}}.
\newblock
\newblock
\shownote{Accessed: 2024}.


\bibitem[Github(2018)]%
        {gitlab}
\bibfield{author}{\bibinfo{person}{Github}.} \bibinfo{year}{2018}\natexlab{}.
\newblock \bibinfo{howpublished}{\url{https://gitlab.com}}.
\newblock
\newblock
\shownote{Accessed: 2024-09-03}.


\bibitem[gokadzev({[n.\,d.]})]%
        {musify}
\bibfield{author}{\bibinfo{person}{gokadzev}.}
  \bibinfo{year}{[n.\,d.]}\natexlab{}.
\newblock \bibinfo{title}{Unlock the full potential of music: Stream
  effortlessly with one app!}
\newblock \bibinfo{howpublished}{\url{https://github.com/gokadzev/Musify}}.
\newblock
\newblock
\shownote{Accessed: 2024-08-14}.


\bibitem[Google(2025)]%
        {android-lint}
\bibfield{author}{\bibinfo{person}{Google}.} \bibinfo{year}{2025}\natexlab{}.
\newblock \bibinfo{title}{Android Lint Rules.}
\newblock
  \bibinfo{howpublished}{\url{https://googlesamples.github.io/android-custom-lint-rules/}}.
\newblock
\newblock
\shownote{Accessed: 2025-02-10}.


\bibitem[Hsiao et~al\mbox{.}(2014)]%
        {hsiao2014race}
\bibfield{author}{\bibinfo{person}{Chun-Hung Hsiao}, \bibinfo{person}{Jie Yu},
  \bibinfo{person}{Satish Narayanasamy}, \bibinfo{person}{Ziyun Kong},
  \bibinfo{person}{Cristiano~L Pereira}, \bibinfo{person}{Gilles~A Pokam},
  \bibinfo{person}{Peter~M Chen}, {and} \bibinfo{person}{Jason Flinn}.}
  \bibinfo{year}{2014}\natexlab{}.
\newblock \showarticletitle{Race detection for event-driven mobile
  applications}.
\newblock \bibinfo{journal}{\emph{ACM SIGPLAN Notices}} \bibinfo{volume}{49},
  \bibinfo{number}{6} (\bibinfo{year}{2014}), \bibinfo{pages}{326--336}.
\newblock


\bibitem[Hu and Neamtiu(2011)]%
        {hu2011automating}
\bibfield{author}{\bibinfo{person}{Cuixiong Hu} {and} \bibinfo{person}{Iulian
  Neamtiu}.} \bibinfo{year}{2011}\natexlab{}.
\newblock \showarticletitle{Automating GUI testing for Android applications}.
  In \bibinfo{booktitle}{\emph{Proceedings of the 6th International Workshop on
  Automation of Software Test}}. ACM, \bibinfo{pages}{77--83}.
\newblock


\bibitem[Huang et~al\mbox{.}(2018)]%
        {huang2018understanding}
\bibfield{author}{\bibinfo{person}{Huaxun Huang}, \bibinfo{person}{Lili Wei},
  \bibinfo{person}{Yepang Liu}, {and} \bibinfo{person}{Shing-Chi Cheung}.}
  \bibinfo{year}{2018}\natexlab{}.
\newblock \showarticletitle{Understanding and detecting callback compatibility
  issues for android applications}. In \bibinfo{booktitle}{\emph{Proceedings of
  the 33rd ACM/IEEE International Conference on Automated Software
  Engineering}}. \bibinfo{pages}{532--542}.
\newblock


\bibitem[Jellyfin(2020)]%
        {jellyfin-android}
\bibfield{author}{\bibinfo{person}{Jellyfin}.} \bibinfo{year}{2020}\natexlab{}.
\newblock \bibinfo{title}{Android Client for Jellyfin.}
\newblock
  \bibinfo{howpublished}{\url{https://github.com/jellyfin/jellyfin-android/}}.
\newblock
\newblock
\shownote{Accessed: 2024-08-14}.


\bibitem[KRTirtho(2021)]%
        {spotube}
\bibfield{author}{\bibinfo{person}{KRTirtho}.} \bibinfo{year}{2021}\natexlab{}.
\newblock \bibinfo{title}{An open source cross-platform Spotify client}.
\newblock \bibinfo{howpublished}{\url{https://github.com/KRTirtho/spotube}}.
\newblock
\newblock
\shownote{Accessed: 2024-08-14}.


\bibitem[Li et~al\mbox{.}(2018)]%
        {li2018cid}
\bibfield{author}{\bibinfo{person}{Li Li}, \bibinfo{person}{Tegawend{\'e}~F
  Bissyand{\'e}}, \bibinfo{person}{Haoyu Wang}, {and} \bibinfo{person}{Jacques
  Klein}.} \bibinfo{year}{2018}\natexlab{}.
\newblock \showarticletitle{Cid: Automating the detection of api-related
  compatibility issues in android apps}. In
  \bibinfo{booktitle}{\emph{Proceedings of the 27th ACM SIGSOFT International
  Symposium on Software Testing and Analysis}}. \bibinfo{pages}{153--163}.
\newblock


\bibitem[Liang et~al\mbox{.}(2014)]%
        {liang2014caiipa}
\bibfield{author}{\bibinfo{person}{Chieh-Jan~Mike Liang},
  \bibinfo{person}{Nicholas~D Lane}, \bibinfo{person}{Niels Brouwers},
  \bibinfo{person}{Li Zhang}, \bibinfo{person}{B{\"o}rje~F Karlsson},
  \bibinfo{person}{Hao Liu}, \bibinfo{person}{Yan Liu}, \bibinfo{person}{Jun
  Tang}, \bibinfo{person}{Xiang Shan}, \bibinfo{person}{Ranveer Chandra},
  {et~al\mbox{.}}} \bibinfo{year}{2014}\natexlab{}.
\newblock \showarticletitle{Caiipa: Automated large-scale mobile app testing
  through contextual fuzzing}. In \bibinfo{booktitle}{\emph{Proceedings of the
  20th annual international conference on Mobile computing and networking}}.
  ACM, \bibinfo{pages}{519--530}.
\newblock


\bibitem[Lillack et~al\mbox{.}(2014)]%
        {lillack2014tracking}
\bibfield{author}{\bibinfo{person}{Max Lillack}, \bibinfo{person}{Christian
  K{\"a}stner}, {and} \bibinfo{person}{Eric Bodden}.}
  \bibinfo{year}{2014}\natexlab{}.
\newblock \showarticletitle{Tracking load-time configuration options}. In
  \bibinfo{booktitle}{\emph{Proceedings of the 29th ACM/IEEE international
  conference on Automated software engineering}}. ACM,
  \bibinfo{pages}{445--456}.
\newblock


\bibitem[Listen-moe(2021)]%
        {listen-moe}
\bibfield{author}{\bibinfo{person}{Listen-moe}.}
  \bibinfo{year}{2021}\natexlab{}.
\newblock \bibinfo{title}{Official LISTEN.moe Android app}.
\newblock
  \bibinfo{howpublished}{\url{https://github.com/LISTEN-moe/android-app}}.
\newblock
\newblock
\shownote{Accessed: 2024-08-14}.


\bibitem[Liu et~al\mbox{.}(2021)]%
        {liu2021identifying}
\bibfield{author}{\bibinfo{person}{Pei Liu}, \bibinfo{person}{Li Li},
  \bibinfo{person}{Yichun Yan}, \bibinfo{person}{Mattia Fazzini}, {and}
  \bibinfo{person}{John Grundy}.} \bibinfo{year}{2021}\natexlab{}.
\newblock \showarticletitle{Identifying and characterizing silently-evolved
  methods in the android API}. In \bibinfo{booktitle}{\emph{2021 IEEE/ACM 43rd
  International Conference on Software Engineering: Software Engineering in
  Practice (ICSE-SEIP)}}. IEEE, \bibinfo{pages}{308--317}.
\newblock


\bibitem[Liu et~al\mbox{.}(2014)]%
        {liu2014characterizing}
\bibfield{author}{\bibinfo{person}{Yepang Liu}, \bibinfo{person}{Chang Xu},
  {and} \bibinfo{person}{Shing-Chi Cheung}.} \bibinfo{year}{2014}\natexlab{}.
\newblock \showarticletitle{Characterizing and detecting performance bugs for
  smartphone applications}. In \bibinfo{booktitle}{\emph{Proceedings of the
  36th International Conference on Software Engineering}}. ACM,
  \bibinfo{pages}{1013--1024}.
\newblock


\bibitem[Maiya et~al\mbox{.}(2014)]%
        {maiya2014race}
\bibfield{author}{\bibinfo{person}{Pallavi Maiya}, \bibinfo{person}{Aditya
  Kanade}, {and} \bibinfo{person}{Rupak Majumdar}.}
  \bibinfo{year}{2014}\natexlab{}.
\newblock \showarticletitle{Race detection for android applications}. In
  \bibinfo{booktitle}{\emph{ACM SIGPLAN Notices}}, Vol.~\bibinfo{volume}{49}.
  ACM, \bibinfo{pages}{316--325}.
\newblock


\bibitem[Mandal et~al\mbox{.}(2018)]%
        {mandal2018vulnerability}
\bibfield{author}{\bibinfo{person}{Amit~Kr Mandal}, \bibinfo{person}{Agostino
  Cortesi}, \bibinfo{person}{Pietro Ferrara}, \bibinfo{person}{Federica
  Panarotto}, {and} \bibinfo{person}{Fausto Spoto}.}
  \bibinfo{year}{2018}\natexlab{}.
\newblock \showarticletitle{Vulnerability analysis of android auto infotainment
  apps}. In \bibinfo{booktitle}{\emph{Proceedings of the 15th ACM International
  Conference on Computing Frontiers}}. \bibinfo{pages}{183--190}.
\newblock


\bibitem[Mandal et~al\mbox{.}(2019)]%
        {mandal2019static}
\bibfield{author}{\bibinfo{person}{Amit~Kr Mandal}, \bibinfo{person}{Federica
  Panarotto}, \bibinfo{person}{Agostino Cortesi}, \bibinfo{person}{Pietro
  Ferrara}, {and} \bibinfo{person}{Fausto Spoto}.}
  \bibinfo{year}{2019}\natexlab{}.
\newblock \showarticletitle{Static analysis of Android Auto infotainment and
  on-board diagnostics II apps}.
\newblock \bibinfo{journal}{\emph{Software: Practice and Experience}}
  \bibinfo{volume}{49}, \bibinfo{number}{7} (\bibinfo{year}{2019}),
  \bibinfo{pages}{1131--1161}.
\newblock


\bibitem[Muccini et~al\mbox{.}(2012)]%
        {muccini2012software}
\bibfield{author}{\bibinfo{person}{Henry Muccini}, \bibinfo{person}{Antonio
  Di~Francesco}, {and} \bibinfo{person}{Patrizio Esposito}.}
  \bibinfo{year}{2012}\natexlab{}.
\newblock \showarticletitle{Software testing of mobile applications: Challenges
  and future research directions}. In \bibinfo{booktitle}{\emph{Proceedings of
  the 7th International Workshop on Automation of Software Test}}. IEEE Press,
  \bibinfo{pages}{29--35}.
\newblock


\bibitem[Namidaco({[n.\,d.]})]%
        {namida}
\bibfield{author}{\bibinfo{person}{Namidaco}.}
  \bibinfo{year}{[n.\,d.]}\natexlab{}.
\newblock \bibinfo{title}{A Beautiful and Feature-rich Music \& Video Player
  with Youtube Support, Built in Flutter}.
\newblock \bibinfo{howpublished}{\url{https://github.com/namidaco/namida}}.
\newblock
\newblock
\shownote{Accessed: 2024-08-14}.


\bibitem[Nextcloud(2013)]%
        {news-android}
\bibfield{author}{\bibinfo{person}{Nextcloud}.}
  \bibinfo{year}{2013}\natexlab{}.
\newblock \bibinfo{title}{Android client for the Nextcloud news/feed reader
  app}.
\newblock
  \bibinfo{howpublished}{\url{https://github.com/nextcloud/news-android}}.
\newblock
\newblock
\shownote{Accessed: 2024-08-14}.


\bibitem[nt4f04uNd({[n.\,d.]})]%
        {sweyer}
\bibfield{author}{\bibinfo{person}{nt4f04uNd}.}
  \bibinfo{year}{[n.\,d.]}\natexlab{}.
\newblock \bibinfo{title}{Music player built with Flutter}.
\newblock \bibinfo{howpublished}{\url{https://github.com/nt4f04uNd/sweyer}}.
\newblock
\newblock
\shownote{Accessed: 2024-08-14}.


\bibitem[OxygenCobalt(2020)]%
        {auxio}
\bibfield{author}{\bibinfo{person}{OxygenCobalt}.}
  \bibinfo{year}{2020}\natexlab{}.
\newblock \bibinfo{title}{A simple, rational music player for android}.
\newblock \bibinfo{howpublished}{\url{https://github.com/OxygenCobalt/Auxio}}.
\newblock
\newblock
\shownote{Accessed: 2024-08-14}.


\bibitem[Quran({[n.\,d.]})]%
        {quran-android}
\bibfield{author}{\bibinfo{person}{Quran}.}
  \bibinfo{year}{[n.\,d.]}\natexlab{}.
\newblock \bibinfo{title}{A quran reading application for android}.
\newblock
  \bibinfo{howpublished}{\url{https://github.com/quran/quran\_android}}.
\newblock
\newblock
\shownote{Accessed: 2024-08-14}.


\bibitem[Rahaman et~al\mbox{.}(2023)]%
        {ast2023}
\bibfield{author}{\bibinfo{person}{Sydur Rahaman}, \bibinfo{person}{Umar
  Farooq}, \bibinfo{person}{Iulian Neamtiu}, {and} \bibinfo{person}{Zhijia
  Zhao}.} \bibinfo{year}{2023}\natexlab{}.
\newblock \showarticletitle{Detecting Potential User-data Save \& Export Losses
  due to Android App Termination}. In \bibinfo{booktitle}{\emph{2023 IEEE/ACM
  International Conference on Automation of Software Test (AST)}}.
  \bibinfo{pages}{152--162}.
\newblock
\urldef\tempurl%
\url{https://doi.org/10.1109/AST58925.2023.00019}
\showDOI{\tempurl}


\bibitem[Scalabrino et~al\mbox{.}(2019)]%
        {scalabrino2019data}
\bibfield{author}{\bibinfo{person}{Simone Scalabrino},
  \bibinfo{person}{Gabriele Bavota}, \bibinfo{person}{Mario
  Linares-V{\'a}squez}, \bibinfo{person}{Michele Lanza}, {and}
  \bibinfo{person}{Rocco Oliveto}.} \bibinfo{year}{2019}\natexlab{}.
\newblock \showarticletitle{Data-driven solutions to detect api compatibility
  issues in android: an empirical study}. In \bibinfo{booktitle}{\emph{2019
  IEEE/ACM 16th International Conference on Mining Software Repositories
  (MSR)}}. IEEE, \bibinfo{pages}{288--298}.
\newblock


\bibitem[Schoon(2024)]%
        {android-auto-number-of-cars}
\bibfield{author}{\bibinfo{person}{Ben Schoon}.}
  \bibinfo{year}{2024}\natexlab{}.
\newblock \bibinfo{title}{Over 200 million cars have Android Auto, a decade
  after its debut}.
\newblock
  \bibinfo{howpublished}{\url{https://9to5google.com/2024/05/16/android-auto-number-of-cars-2024/}}.
\newblock
\newblock
\shownote{Accessed: 2024-09-03}.


\bibitem[SimpleMobileTools({[n.\,d.]})]%
        {simplemusicplayer}
\bibfield{author}{\bibinfo{person}{SimpleMobileTools}.}
  \bibinfo{year}{[n.\,d.]}\natexlab{}.
\newblock \bibinfo{title}{A clean music player with a customizable widget,
  stylish interface and no ads.}
\newblock
  \bibinfo{howpublished}{\url{https://github.com/SimpleMobileTools/Simple-Music-Player}}.
\newblock
\newblock
\shownote{Accessed: 2024-08-14}.


\bibitem[Sosauce({[n.\,d.]})]%
        {cutemusic}
\bibfield{author}{\bibinfo{person}{Sosauce}.}
  \bibinfo{year}{[n.\,d.]}\natexlab{}.
\newblock \bibinfo{title}{CuteMusic is a simple,lightweight and open-source
  offline music player app for Android.}
\newblock \bibinfo{howpublished}{\url{https://github.com/sosauce/CuteMusic}}.
\newblock
\newblock
\shownote{Accessed: 2024-08-14}.


\bibitem[Timusus({[n.\,d.]})]%
        {shuttle2}
\bibfield{author}{\bibinfo{person}{Timusus}.}
  \bibinfo{year}{[n.\,d.]}\natexlab{}.
\newblock \bibinfo{title}{Shuttle Music Player 2.0}.
\newblock \bibinfo{howpublished}{\url{https://github.com/timusus/Shuttle2}}.
\newblock
\newblock
\shownote{Accessed: 2024-08-14}.


\bibitem[Wang et~al\mbox{.}(2019)]%
        {wang2019characterizing}
\bibfield{author}{\bibinfo{person}{Haoyu Wang}, \bibinfo{person}{Hongxuan Liu},
  \bibinfo{person}{Xusheng Xiao}, \bibinfo{person}{Guozhu Meng}, {and}
  \bibinfo{person}{Yao Guo}.} \bibinfo{year}{2019}\natexlab{}.
\newblock \showarticletitle{Characterizing android app signing issues}. In
  \bibinfo{booktitle}{\emph{2019 34th IEEE/ACM International Conference on
  Automated Software Engineering (ASE)}}. IEEE, \bibinfo{pages}{280--292}.
\newblock


\bibitem[Wei et~al\mbox{.}(2016)]%
        {wei2016taming}
\bibfield{author}{\bibinfo{person}{Lili Wei}, \bibinfo{person}{Yepang Liu},
  {and} \bibinfo{person}{Shing-Chi Cheung}.} \bibinfo{year}{2016}\natexlab{}.
\newblock \showarticletitle{Taming android fragmentation: Characterizing and
  detecting compatibility issues for android apps}. In
  \bibinfo{booktitle}{\emph{Proceedings of the 31st IEEE/ACM International
  Conference on Automated Software Engineering}}. \bibinfo{pages}{226--237}.
\newblock


\bibitem[Wei et~al\mbox{.}(2018)]%
        {wei2018understanding}
\bibfield{author}{\bibinfo{person}{Lili Wei}, \bibinfo{person}{Yepang Liu},
  \bibinfo{person}{Shing-Chi Cheung}, \bibinfo{person}{Huaxun Huang},
  \bibinfo{person}{Xuan Lu}, {and} \bibinfo{person}{Xuanzhe Liu}.}
  \bibinfo{year}{2018}\natexlab{}.
\newblock \showarticletitle{Understanding and detecting fragmentation-induced
  compatibility issues for android apps}.
\newblock \bibinfo{journal}{\emph{IEEE Transactions on Software Engineering}}
  \bibinfo{volume}{46}, \bibinfo{number}{11} (\bibinfo{year}{2018}),
  \bibinfo{pages}{1176--1199}.
\newblock


\bibitem[Yan et~al\mbox{.}(2013)]%
        {yan2013systematic}
\bibfield{author}{\bibinfo{person}{Dacong Yan}, \bibinfo{person}{Shengqian
  Yang}, {and} \bibinfo{person}{Atanas Rountev}.}
  \bibinfo{year}{2013}\natexlab{}.
\newblock \showarticletitle{Systematic testing for resource leaks in Android
  applications}. In \bibinfo{booktitle}{\emph{Software Reliability Engineering
  (ISSRE), 2013 IEEE 24th International Symposium on}}. IEEE,
  \bibinfo{pages}{411--420}.
\newblock


\bibitem[Zhang et~al\mbox{.}(2019)]%
        {auto-messages-testing}
\bibfield{author}{\bibinfo{person}{Yu Zhang}, \bibinfo{person}{Xi Deng},
  \bibinfo{person}{Jun Yan}, \bibinfo{person}{Hang Su}, {and}
  \bibinfo{person}{Hongyu Gao}.} \bibinfo{year}{2019}\natexlab{}.
\newblock \showarticletitle{Testing the Message Flow of Android Auto Apps}. In
  \bibinfo{booktitle}{\emph{2019 IEEE 26th International Conference on Software
  Analysis, Evolution and Reengineering (SANER)}}. \bibinfo{pages}{559--563}.
\newblock
\urldef\tempurl%
\url{https://doi.org/10.1109/SANER.2019.8667973}
\showDOI{\tempurl}


\end{thebibliography}

\end{document}